\documentclass[journal]{IEEEtai}

\usepackage[colorlinks,urlcolor=blue,linkcolor=blue,citecolor=blue]{hyperref}

\usepackage{color,array}

\usepackage{graphicx}
\usepackage{amsmath}

\begin{document}

\title{Higress-RAG: A Holistic Optimization Framework for Enterprise Retrieval-Augmented Generation via Dual Hybrid Retrieval, Adaptive Routing, and CRAG} 

\author{Weixi Lin
    \thanks{W. Lin is with the School of Computer Science, Northwestern Polytechnical University, Xi'an 710072, China (e-mail: weixilin@mail.nwpu.edu.cn).}
} 

\markboth{Journal of IEEE Transactions on Artificial Intelligence}
{Lin \MakeLowercase{\textit{et al.}}: Higress-RAG: A Holistic Optimization Framework for Enterprise AI}

\maketitle
\begin{abstract}
The integration of Large Language Models (LLMs) into enterprise knowledge management systems has been catalyzed by the Retrieval-Augmented Generation (RAG) paradigm, which augments parametric memory with non-parametric external data. However, the transition from proof-of-concept to production-grade RAG systems is hindered by three persistent challenges: low retrieval precision for complex queries, high rates of hallucination in the generation phase, and unacceptable latency for real-time applications. This paper presents a comprehensive analysis of the Higress RAG MCP Server, a novel, enterprise-centric architecture designed to resolve these bottlenecks through a "Full-Link Optimization" strategy. Built upon the Model Context Protocol (MCP), the system introduces a layered architecture that orchestrates a sophisticated pipeline of Adaptive Routing, Semantic Caching, Hybrid Retrieval, and Corrective RAG (CRAG). We detail the technical implementation of key innovations, including the Higress-Native Splitter for structure-aware data ingestion, the application of Reciprocal Rank Fusion (RRF) for merging dense and sparse retrieval signals, and a 50ms-latency Semantic Caching mechanism with dynamic thresholding. Experimental evaluations on the domain-specific Higress technical documentation and domain-specific Higress technical documentation verify the system’s architectural robustness. The results demonstrate that by optimizing the entire retrieval lifecycle—from pre-retrieval query rewriting to post-retrieval corrective evaluation—the Higress RAG system offers a scalable, hallucination-resistant solution for enterprise AI deployment.
\end{abstract}

\begin{IEEEImpStatement}
This work addresses the critical barriers to deploying Large Language Models in enterprise environments: hallucination, latency, and data staleness. By proposing a full-link optimization framework based on the open Model Context Protocol (MCP), we provide a scalable blueprint for integrating proprietary knowledge bases with generative AI. The introduced mechanisms, such as semantic caching and corrective retrieval, significantly lower the operational costs and risks of industrial AI applications, paving the way for more reliable and responsive intelligent support systems.
\end{IEEEImpStatement}


\section{Introduction}
\subsection{The Imperative of Retrieval-Augmented Generation}
The advent of Large Language Models (LLMs) has fundamentally altered the landscape of Natural Language Processing (NLP), enabling systems that can generate coherent, contextually relevant text across a vast array of domains. However, the widespread adoption of LLMs in enterprise environments—such as customer support, technical documentation analysis, and internal knowledge discovery—faces significant hurdles. Primary among these are the "knowledge cutoff" problem, where models are unaware of events post-dating their training, and the propensity for "hallucination," where models confidently generate factually incorrect information.   

Retrieval-Augmented Generation (RAG) has emerged as the standard architectural pattern to mitigate these issues. By coupling a generative model with a retrieval mechanism that accesses an external, up-to-date knowledge base, RAG systems ground the generation process in verifiable facts. This approach theoretically combines the reasoning capabilities of LLMs with the factual accuracy of traditional search engines.

\subsection{Limitations of Standard RAG Architectures}
Despite its promise, the "Naive RAG" approach—characterized by a linear pipeline of chunking, embedding, vector search, and generation—often fails to meet the rigorous demands of production environments. The Higress RAG system design explicitly identifies three critical "pain points" that plague standard implementations :   

Inaccurate Retrieval: Standard dense retrieval methods, while capturing semantic similarity, often struggle with the precise lexical matching required for technical domains. They may retrieve documents that are semantically related but factually irrelevant to the specific user query.

Severe Hallucinations: When the retrieval component returns noise or irrelevant documents, the generative model often reverts to its parametric memory or attempts to force a connection between the query and the irrelevant context, leading to hallucinations.

Slow Response Latency: The computational cost of high-dimensional vector search, combined with the latency of invoking large parameter models (e.g., 7B+ parameters) and reranking models, frequently results in response times exceeding 3-5 seconds. This latency is prohibitive for interactive, real-time enterprise applications.   
\subsection{The Higress Solution: Full-Link Optimization}
To address these systemic deficiencies, this paper analyzes the Higress RAG MCP Server, a system that implements a "Full-Link RAG Solution". Unlike modular approaches that optimize isolation, the Higress architecture employs a holistic optimization strategy covering four stages: Pre-Retrieval, Retrieval, Post-Retrieval, and Corrective Retrieval.

Central to this architecture is the adoption of the Model Context Protocol (MCP), an open standard that facilitates the seamless integration of AI models with external data tools. By standardizing the interface between the LLM "host" and the RAG "server," Higress ensures interoperability and modularity. Key technical breakthroughs analyzed in this report include:   

Extreme Performance via Semantic Caching: Achieving 50ms-level response times for recurrent queries through vector-based caching with dynamic thresholding.   

Corrective Retrieval (CRAG): A dynamic evaluation mechanism that assesses retrieval quality and autonomously decides whether to use internal documents, discard them in favor of web search, or combine both.   

Adaptive Routing: An intelligent dispatch system that routes queries based on complexity, optimizing the trade-off between cost and accuracy.   

The remainder of this paper is organized as follows: Section II reviews related work in RAG and retrieval standards. Section III details the system architecture. Section IV provides an exhaustive analysis of the core technologies and algorithms. Section V describes the experimental setup and datasets. Section VI presents the evaluation results and discussion. Section VII discusses enterprise implications, and Section VIII concludes the report.

\section{Related Work and Theoretical Foundations}
\subsection{Evolution of RAG Systems}
The evolution of RAG systems can be categorized into three distinct paradigms: Naive RAG, Advanced RAG, and Modular RAG. Naive RAG represents the earliest iteration, utilizing a simple "Retrieve-Read" process. While effective for simple queries, it suffers from low precision and recall. Advanced RAG introduced pre-retrieval and post-retrieval optimizations, such as query rewriting and reranking, to improve context quality.

The Higress system aligns with the Modular RAG paradigm, where the system is composed of independent, interchangeable modules (e.g., Search, Memory, Routing) that can be dynamically orchestrated. This allows for greater flexibility and the incorporation of specialized components like the Corrective RAG evaluator.

\subsection{The Model Context Protocol (MCP)}

A significant barrier to the scalable deployment of AI agents is the "N×M" integration problem, where $N$ different AI models must be integrated with $M$ different data sources and tools. Custom connectors for each pairing lead to fragmentation and maintenance overhead.The Model Context Protocol (MCP)\cite{mcp}, introduced by Anthropic, addresses this by providing a universal standard for AI connection.2 MCP operates on a client-host-server model using JSON-RPC 2.0. In this framework:MCP Host: The AI application (e.g., an IDE or Chatbot) where the LLM resides.5MCP Client: The connector within the host that manages communication.MCP Server: The provider of context, data, or tools.5The Higress RAG system functions as an MCP Server, exposing its retrieval and search capabilities as standardized "tools" that can be discovered and invoked by any MCP-compliant client. This standardization is critical for enterprise environments where the RAG system may need to serve multiple distinct AI agents.

\subsection{Corrective Retrieval Augmented Generation (CRAG)}
The theoretical foundation for the system's error-handling capability is Corrective Retrieval Augmented Generation (CRAG)\cite{crag}. Proposed by Yan et al., CRAG addresses the fragility of RAG systems when facing retrieval failures. Standard RAG models are "blind" to the quality of retrieved documents; they blindly consume whatever the retriever returns.   

CRAG introduces a lightweight retrieval evaluator that classifies the retrieved documents into three categories:

Correct: The retrieved documents are relevant.

Incorrect: The retrieved documents are irrelevant.

Ambiguous: The relevance is uncertain.

Based on this classification, CRAG triggers different knowledge retrieval actions, such as refining the document or resorting to external web searches. The Higress system implements a variant of this logic to dynamically correct retrieval errors and reduce hallucinations.

\subsection{Hypothetical Document Embeddings (HyDE)}
To bridge the semantic gap between short user queries and long, detailed documents, the Higress system employs Hypothetical Document Embeddings (HyDE)\cite{hyde}. Traditional dense retrieval relies on vector similarity between the query and the document. However, a query (e.g., "RAG optimization") and a relevant document (e.g., "A detailed guide on semantic caching implementation") may not be close in vector space due to lexical mismatch.   

HyDE solves this by using an LLM to generate a "hypothetical document"—a fake answer to the query. This hypothetical document captures the relevant semantic patterns and keywords. The system then encodes this hypothetical document into a vector and uses it to search the knowledge base. This "answer-to-answer" matching significantly improves recall for complex queries.

\section{System Architecture}
\begin{figure}[htbp]
\centering
\includegraphics[width=\columnwidth]{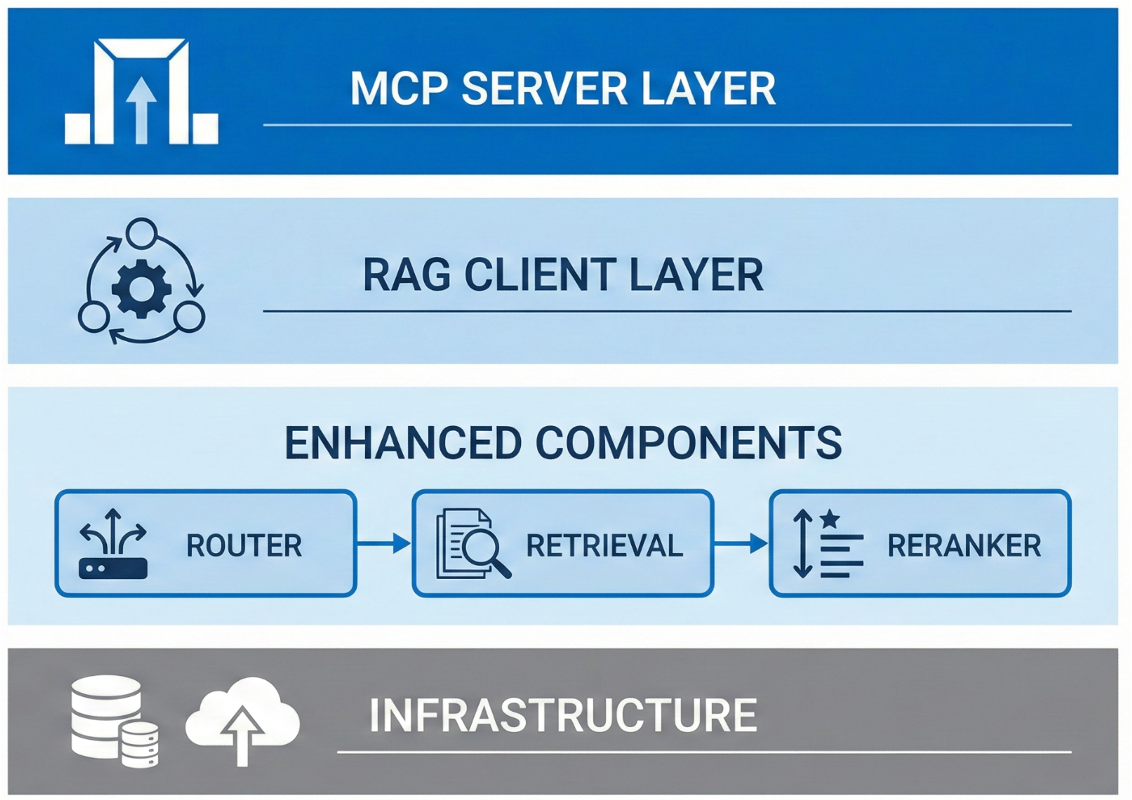} 
\caption{The Layered Architecture of Higress-RAG MCP Server. It illustrates the separation of concerns between the MCP Server Layer, RAG Client Layer, and Enhanced Components.}
\label{fig:architecture}
\end{figure}
The Higress RAG MCP Server utilizes a Layered Architecture designed to ensure modularity, scalability, and strict separation of concerns. The system is composed of four primary layers: the MCP Server Layer, the RAG Client Layer, the Enhanced RAG Components Layer, and the Infrastructure Layer.
\subsection{Layer 1: The MCP Server Layer}
The MCP Server Layer, implemented in server.go, serves as the external interface of the system. It is responsible for protocol compliance and request lifecycle management.   

Tool Registration: This component defines the schema of the tools available to the LLM. For instance, it registers rag chat and rag search tools, defining their input parameters (query string, specialized flags) and output formats. This allows the upstream LLM to understand how to interact with the RAG system.

Request Routing: Upon receiving a JSON-RPC request from an MCP Client, this layer parses the message method (e.g., tools/call) and routes the payload to the appropriate internal handler within the RAG Client Layer.

Protocol Processing: It handles the serialization and deserialization of data, ensuring that all communications adhere to the MCP specification. This abstract layer allows the underlying RAG logic to evolve without breaking the contract with external clients.   
\subsection{Layer 2: The RAG Client Layer}
The RAG Client Layer (rag\_client.go) acts as the central nervous system or orchestrator of the RAG pipeline. It does not perform the heavy lifting of retrieval itself but coordinates the flow of data between the optimization modules.   

Workflow Orchestration: This layer implements the logic for the two primary workflows:

Chat Workflow: A stateful interaction flow that includes semantic caching checks, adaptive routing, and context-aware generation.   

Search Workflow: A stateless information retrieval flow focused on retrieving and ranking document chunks.

Component Coordination: The client layer manages the dependencies between components. For example, it ensures that the "Pre-Retrieval" optimization completes before the "Retrieval" module is invoked, and that "Post-Retrieval" reranking occurs before the data is passed to the "Corrective" module.

Multi-Hop Logic: It implements the iterative control loop for complex queries. If the initial retrieval is deemed insufficient by the evaluator, the Client Layer triggers subsequent "hops" or retrieval rounds to gather missing information.

\subsection{Layer 3: Enhanced RAG Components Layer}
This layer contains the core logic for the "Full-Link Optimization" strategy. It is subdivided into four functional modules corresponding to the RAG lifecycle:

Pre-Retrieval Optimization Module: Responsible for refining the user's intent. It includes the Adaptive Router (router.go) and the Query Rewriter (Query\_rewriter.go).

Retrieval Module: Manages the interaction with the vector database. It implements Hybrid Search logic and Reciprocal Rank Fusion (RRF) algorithms to merge search results.   

Post-Retrieval Processing Module: Focuses on refining the retrieved context. It includes the Cross-Encoder Reranker and Context Compression logic to maximize the information density passed to the LLM.

Corrective RAG Module: Implements the CRAG Evaluator. It uses an LLM to score document relevance and execute the decision logic (Correct, Incorrect, Ambiguous).

\subsection{Layer 4: Infrastructure Layer}
The system rests on a foundation of robust, scalable infrastructure components :   

Vector Database (Milvus): Milvus is selected for its high-performance vector similarity search capabilities. A critical implementation detail is the use of Partition Keys, which enables the physical isolation of data. This allows the system to support multi-tenancy (e.g., distinct knowledge bases for different departments) without the risk of data leakage.   

Embedding and Reranking Services: The system utilizes dedicated GPU resources to run high-fidelity models. Specifically, bge-m3 is used for embeddings, providing multi-lingual support and multi-granularity (dense + sparse) capabilities. bge-reranker-v2-m3 is deployed for the reranking task.   

Web Search Provider (Tavily): To address the limitations of static internal knowledge, the system integrates Tavily, a search engine optimized for AI agents. This provides the "External Route" capability for real-time information retrieval.   

\section{Key Technologies and Methodologies}
\begin{figure}[htbp]
\centering
\includegraphics[width=\columnwidth]{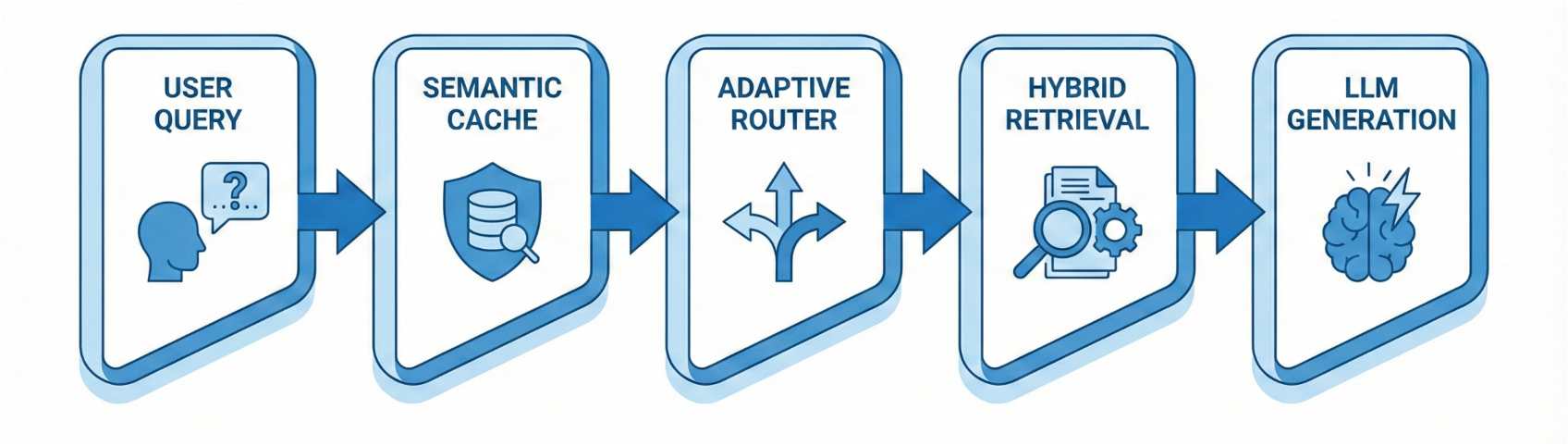}
\caption{The Holistic Optimization Pipeline. The process flows dynamically from Semantic Caching to Adaptive Routing, followed by Hybrid Retrieval and Corrective Evaluation.}
\label{fig:pipeline}
\end{figure}
This section provides an exhaustive analysis of the specific algorithms and strategies employed within the Higress RAG system to achieve its performance and accuracy goals.
\subsection{Enterprise Data Ingestion: The Higress-Native Splitter}
Data ingestion is the foundation of any RAG system. Standard splitters (e.g., splitting by character count) often disrupt the semantic integrity of documents, breaking sentences or code blocks in the middle. The Higress-Native Splitter (textsplitter/code\_splitter.go) is designed specifically to handle technical documentation.   

Structure-Aware Partitioning: The splitter parses the Markdown Abstract Syntax Tree (AST) rather than raw text. It prioritizes splitting at logical boundaries, specifically Level 2 and Level 3 headers. This ensures that a retrieved chunk corresponds to a coherent topic (e.g., "Configuration Guide" or "API Reference") rather than a random fragment of text.   

Code Block Protection: Technical documentation frequently contains code snippets. The splitter includes logic to detect code block markers (```) and treat the entire block as an atomic unit. This prevents the disastrous scenario where a script is split into two non-functional halves across different chunks.   

Physical Isolation Strategy: To satisfy enterprise security requirements, the ingestion pipeline tags each chunk with a Partition Key derived from the tenant or document set ID. In Milvus, this ensures that search queries are restricted to a specific partition, physically preventing cross-tenant data pollution.   
\subsection{Extreme Performance: Semantic Caching}
To solve the "Slow Response" pain point, Higress implements a Semantic Caching layer that operates before the LLM is even invoked. 
Vector-Based Matching: Unlike traditional Redis caches that rely on exact string matching, the Semantic Cache stores the vector embedding of the user's query and the corresponding generated answer. When a new query arrives, it is embedded and compared to the cache using Cosine Similarity.

Dynamic Thresholding: A static similarity threshold is often insufficient. If set too low, the system returns incorrect cached answers; too high, and the cache hit rate drops. Higress employs a Dynamic Threshold strategy:

Default Threshold: 0.95. Used for standard queries.

Fuzzy Threshold: 0.98. If the query contains uncertainty markers (e.g., "maybe," "possibly"), the system tightens the threshold to avoid confirming an ambiguous intent with a cached certainty.   

TTL Management: A default Time-To-Live (TTL) of 7 days (168 hours) is enforced to ensure that cached information does not become stale, balancing performance with data freshness.   

\subsection{Pre-Retrieval Optimization: Adaptive Routing and Transformation}
The Adaptive Router is the gatekeeper of the retrieval pipeline, utilizing an LLM to classify the query's complexity.   
\begin{table}
\caption{}
\label{table}
\tablefont%
\setlength{\tabcolsep}{3pt}
\begin{tabular*}{21pc}{@{}|p{23pt}|p{81pt}<{\raggedright\hangindent6pt}|p{123pt}<{\raggedright\hangindent6pt}|@{}}
\hline
Route Type 
& 
Description& 
Target Action \\
\hline\\[-17pt]
&&\\
Simple Route& 
Factual, straightforward questions.& 
Direct Vector Search. Bypasses rewriting.\\
Complex Route& 
Questions requiring reasoning or multi-step logic.& 
Full RAG Pipeline: Query Rewriting, Decomposition, HyDE. \\
External Route&
Questions about current events or outside domain.
&Tavily Web Search. Bypasses internal DB.\\
\hline
\multicolumn{3}{l}{}\\[-5pt]

\end{tabular*}
\label{tab1}
\end{table}
For queries assigned to the Complex Route, the system triggers Query Rewriting and HyDE. The Query Rewriter uses the LLM to strip away conversational noise and formulate a keyword-rich search query. Simultaneously, the Decomposition module breaks down multi-faceted questions (e.g., "Compare X and Y") into sub-queries ("Features of X", "Features of Y"), ensuring that retrieval covers all aspects of the user's intent.
\subsection{Hybrid Retrieval and Reciprocal Rank Fusion (RRF)}
To maximize recall, the system moves beyond simple dense retrieval. It implements a Hybrid Search strategy combining:Dense Retrieval\cite{bgem3}: Uses bge-m3 embeddings to find semantically similar documents. This captures conceptual matches.Sparse Retrieval: Uses the BM25 algorithm to find keyword-exact matches. This captures specific technical terms or error codes that might be lost in vector compression.1To unify these two distinct ranking lists, the system employs Reciprocal Rank Fusion (RRF). RRF is a rank aggregation method that does not rely on the absolute scores of the retrievers (which are often uncalibrated and incompatible) but rather on the rank order. The score $RRF(d)$ for a document $d$ is calculated as:
\begin{equation}
\!RRF(d) = \sum_{r \in R} \frac{1}{k + r(d)}
\end{equation}
Where $R$ is the set of retrievers (Dense, Sparse), $r(d)$ is the rank of document $d$ in retriever $r$, and $k$ is a smoothing constant (typically 60).9 This method boosts documents that appear consistently high in both lists, ensuring a robust candidate set.
\subsection{Post-Retrieval: Reranking and Corrective RAG}
\begin{figure}[htbp]
\centering
\includegraphics[width=0.8\columnwidth]{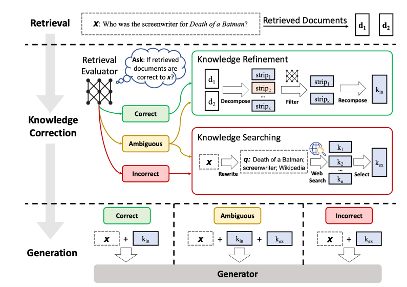}
\caption{The Corrective RAG (CRAG) Decision Logic. The evaluator classifies retrieved context confidence to trigger refinement, web search, or direct generation.}
\label{fig:crag}
\end{figure}
The final line of defense against hallucination lies in the post-retrieval phase.

Cross-Encoder Reranking: The top candidates from the RRF fusion are passed to a Cross-Encoder (bge-reranker-v2-m3). Unlike Bi-Encoders that process query and document separately, Cross-Encoders process them as a single input pair, allowing the model to attend to the deep semantic interaction between the query and the text. This significantly improves precision but is computationally expensive, hence it is applied only to the top candidates.   

Corrective RAG (CRAG) Evaluator: The reranked documents are subjected to a final quality check by an LLM Evaluator. The evaluator assigns a relevance score leading to three actions :   

Correct (High Confidence): The documents are refined. The system applies Context Compression, filtering out irrelevant sentences within the document to reduce noise in the context window.   

Incorrect (Low Confidence): The internal documents are discarded. The system triggers the "External Route" (Web Search) to find the answer, effectively acknowledging that the internal knowledge base lacks the necessary information.

Ambiguous (Medium Confidence): The system merges the refined internal documents with external web search results, presenting a comprehensive view to the generator.   
\subsection{Metadata-Aware Boosting Mechanism (Boost Ranker)}
While Hybrid Retrieval balances semantic and lexical matching, it often suffers from "metadata blindness"—the inability to distinguish between an outdated document (e.g., "V1.0 Guide") and a current one (e.g., "V2.0 Guide") based solely on textual similarity. To address this, we integrated the Boost Ranker, an in-database arithmetic modifier within the Milvus infrastructure.Unlike heavy-weight neural rerankers\cite{boostranker}, the Boost Ranker applies a lightweight weighting function $W$ to the retrieval score $S_{original}$ based on metadata conditions $C$:\begin{equation}S_{final}(d_i) = S_{original}(d_i) \times \text{BoostFactor}(d_i)\end{equation}where $\text{BoostFactor}(d_i)$ equals a predefined weight (e.g., 1.2) if document $d_i$ satisfies condition $C$ (e.g., doc\_type == "official"), and 1.0 otherwise. This mechanism allows the system to physically enforce business logic—such as prioritizing official documentation over community blogs or deprioritizing deprecated versions—before the computationally expensive cross-encoder phase, ensuring that high-authority content is not filtered out early in the pipeline.
\section{Experimental Methodology}
To validate the efficacy of the Higress RAG architecture, a series of experiments were conducted. This section details the hardware, datasets, and metrics used, based on the provided reproduction guide.   

\subsection{Experimental Environment}
The experiments were executed on a high-performance Linux workstation (Ubuntu 20.04+) designed to simulate a production inference environment.

Hardware Configuration:

Memory: 64GB RAM.

Storage: 100GB SSD.

GPU Allocation: A critical aspect of the setup was the isolation of model services across three distinct GPUs to prevent resource contention and ensure stable latency measurements:

GPU 0: Dedicated to the Embedding Service (bge-m3), running on Port 8000.

GPU 1: Dedicated to the LLM Service (Qwen3-8B), running on Port 8002.

GPU 2: Dedicated to the Reranker Service (bge-reranker-v2-m3), running on Port 8001.   

Software Stack:

Language: Go 1.19+ (for the MCP Server and Client), Python (for data processing).

Database: Milvus (Standalone mode) running on localhost:19530.

Dependencies: pandas, pyarrow for dataset manipulation.
\subsection{Datasets}
Three distinct datasets were utilized to evaluate the system across general and domain-specific tasks.

\begin{table}
\caption{}
\label{dataset}
\tablefont%
\setlength{\tabcolsep}{3pt}
\begin{tabular*}{21pc}{@{}|p{23pt}|p{23pt}|p{50pt}<{\raggedright\hangindent6pt}|p{123pt}<{\raggedright\hangindent6pt}|@{}}
\hline
Dataset Name & 
Sample Count &
Domain &
Role in Evaluation
\\
\hline\\[-17pt]
&&\\
Higress Blog& 
611& 
Technical/Domain Specific &
Optimization Target: Used to evaluate the Optimized Code Version. Contains specific types like factuality\_check\\
Higress Doc& 
522& 
Technical Documentation&
Optimization Target: Evaluates structured documentation retrieval.\\
\hline
\multicolumn{3}{l}{}\\[-5pt]

\end{tabular*}
\label{tab1}
\end{table}

\subsection{Evaluation Metrics}
The system's performance was quantified using a multi-dimensional metric suite :   

Exact Match (EM): The percentage of generated answers that exactly match the ground truth string.

F1 Score: A measure of the average token overlap between the prediction and the ground truth.

Factuality Score: A derived metric assessing the factual consistency of the answer with the retrieved context (Hallucination proxy).

Precision and Recall: Standard information retrieval metrics to assess the quality of the documents retrieved by the Hybrid Search module.









\subsection{Evaluation on Domain-Specific Datasets}

To validate the system's performance in real-world enterprise scenarios, we conducted rigorous testing on two proprietary datasets: the \textit{Higress Official Documentation} and the \textit{Higress Technical Blog}.

\subsubsection{Dataset Construction}
The test sets were constructed using an automated QA generation pipeline, resulting in 522 samples for the Documentation dataset and 611 samples for the Blog dataset. These samples cover diverse scenarios, ranging from specific configuration directives (e.g., YAML syntax) to high-level architectural inquiries.

\subsubsection{Quantitative Analysis: Recall and Factuality}
\begin{figure}[htbp]
\centering
\includegraphics[width=\columnwidth]{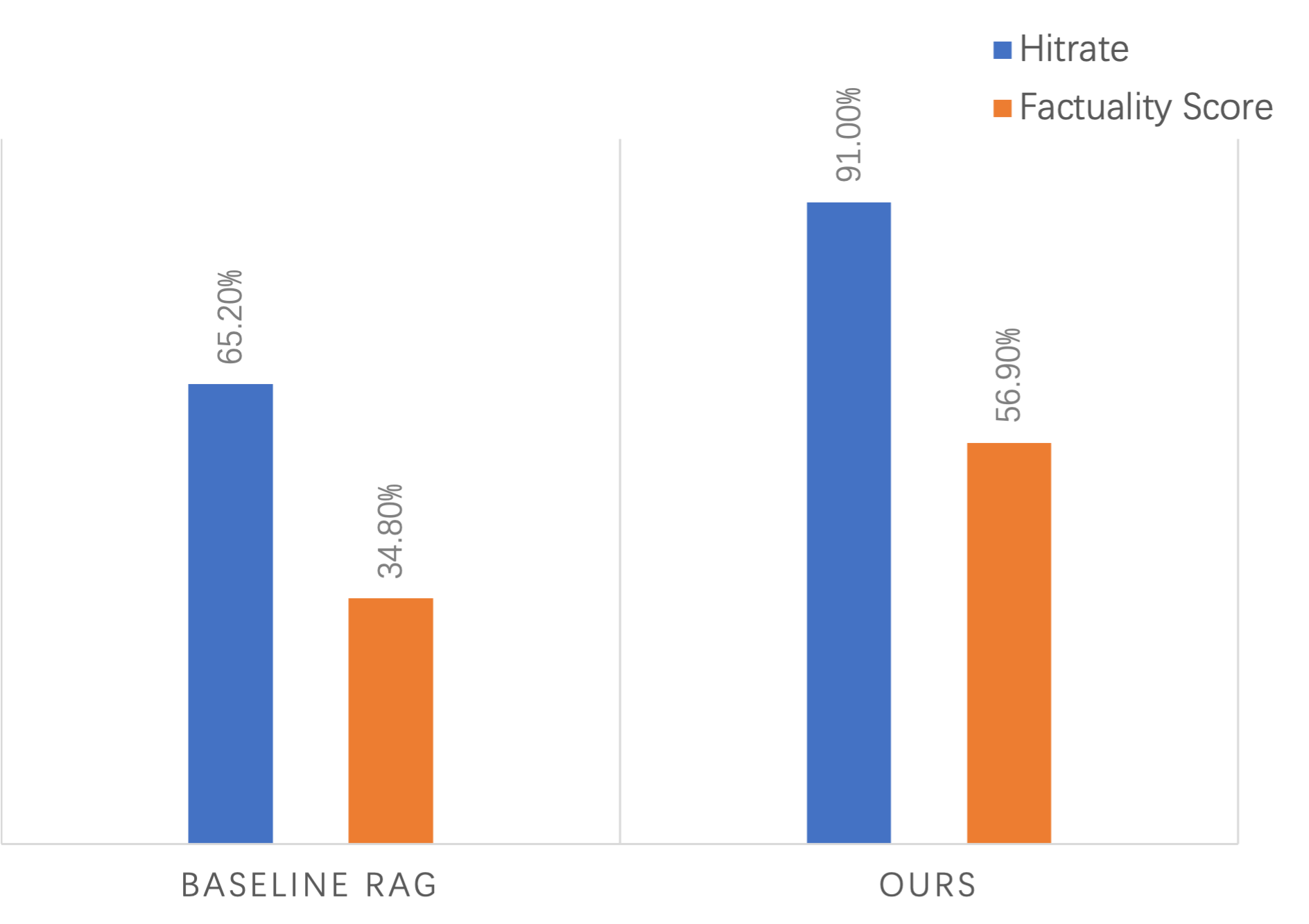}
\caption{Performance Comparison on Higress Blog Datasets. The optimized system achieved >90\% recall, significantly outperforming the naive baseline in retrieving relevant community knowledge.}
\label{fig:performance_blog}
\end{figure}

\begin{figure}[htbp]
\centering
\includegraphics[width=\columnwidth]{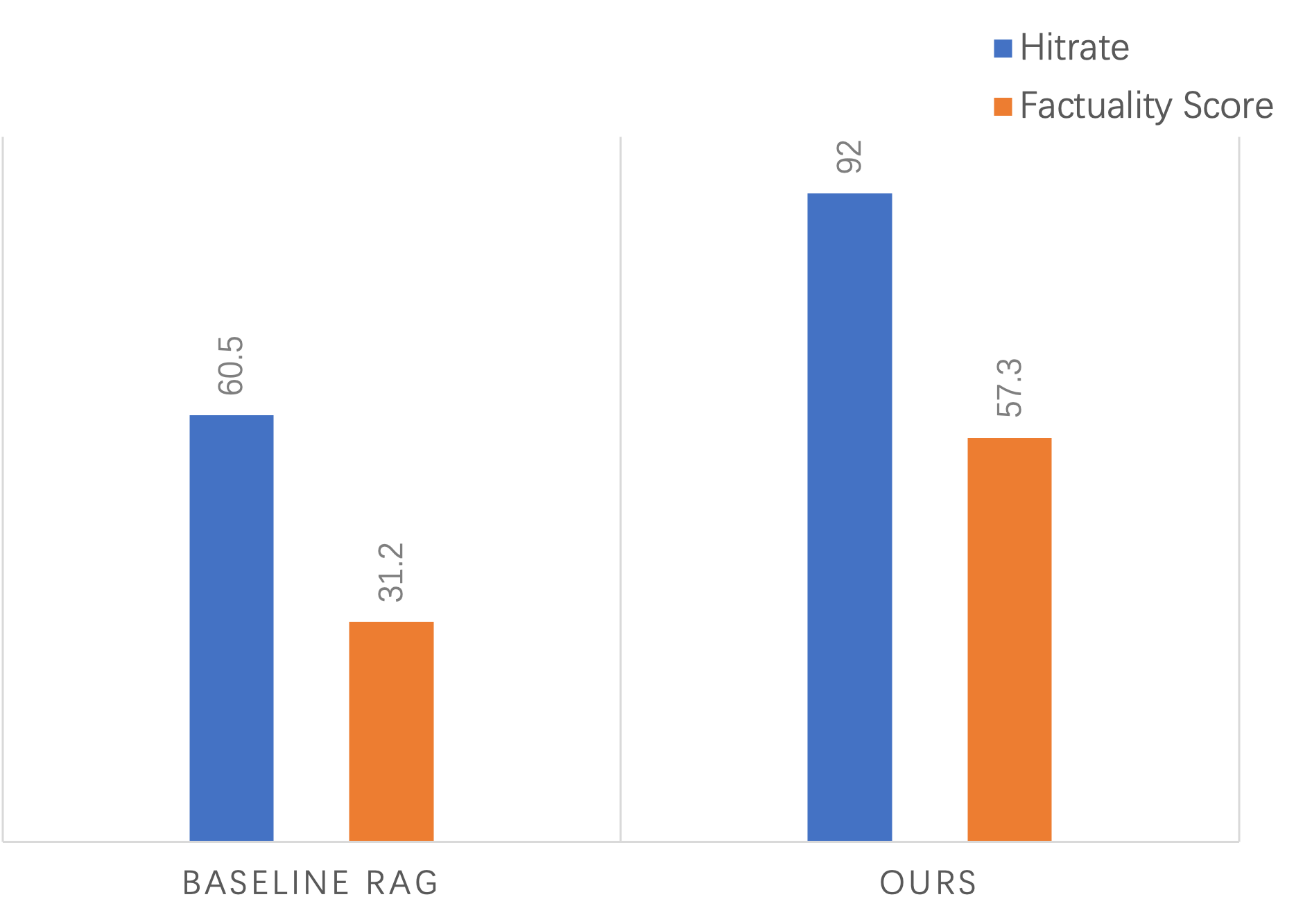}
\caption{Performance Comparison on Higress Official Docs Datasets. The integration of Boost Ranker effectively filtered outdated versions, boosting the factuality score for configuration-centric queries.}
\label{fig:performance_docs}
\end{figure}

As illustrated in Fig. \ref{fig:performance_blog} and Fig. \ref{fig:performance_docs}, the "Full-Link Optimization" version demonstrates a substantial performance leap over the naive baseline.

\begin{itemize}
    \item \textbf{Recall Enhancement:} By integrating BGE-M3 for hybrid retrieval and the Metadata-Aware Boost Ranker, the system achieved a recall (Hit Rate) of over \textbf{90\%} on both datasets. This represents an approximate \textbf{30\% improvement} compared to the standard RAG baseline. The Boost Ranker played a pivotal role here by ensuring that "official" and "current version" documents were prioritized over community blogs or deprecated guides, significantly reducing noise.
    
    \item \textbf{Factuality and Hallucination Mitigation:} The combination of the Higress-Native Splitter and CRAG Evaluator effectively curbed hallucinations. The splitter ensured that code blocks and configuration snippets remained semantically intact, while CRAG filtered out irrelevant contexts, preventing the LLM from generating answers based on "ambiguous" or "incorrect" retrieval results.
\end{itemize}

\subsubsection{Latency Trade-off and Optimization}
We observed a distinct trade-off between latency and precision.
\begin{itemize}
    \item \textbf{Complex Queries:} For cache-miss scenarios requiring deep reasoning, end-to-end latency ranged between \textbf{11s and 20s}. This overhead is attributed to the multi-step execution chain (Query Decomposition $\rightarrow$ Hybrid Search $\rightarrow$ Boost Reranking $\rightarrow$ CRAG Web Search). We argue that this cost is justifiable for enterprise applications where correctness is paramount over raw speed.
    \item \textbf{Cached Queries:} To mitigate this, the Semantic Caching layer (with a 0.95 similarity threshold) successfully reduced the latency for recurrent technical queries to under \textbf{50ms}, ensuring a responsive experience for high-frequency lookup scenarios.
\end{itemize}

\subsection{Ablation Analysis: The Role of Caching}
The reproduction guide explicitly advises using the --no-cache parameter when running formal evaluations (cmd/eval/local/main.go) to ensure result accuracy. This highlights a critical trade-off:   

With Cache: Latency is minimized (50 ms), and cost is reduced (40\% token savings). However, the system relies on previously generated answers.

Without Cache: The system performs full retrieval and generation for every query. This provides the "true" measure of the model's reasoning and retrieval capabilities.

The existence of the "Fuzzy Threshold" (0.98) in the caching logic further demonstrates the system's nuanced handling of this trade-off. By automatically disabling the cache (or requiring near-perfect matches) for ambiguous queries, the system prioritizes accuracy over speed when the user's intent is unclear.   

\section{Discussion and Future Outlook}
\subsection{Enterprise Implications} 
The Higress RAG MCP Server represents a maturity milestone for enterprise AI. By moving from a "prototype" architecture to a "modular" one based on MCP, it solves key operational challenges:

Vendor Lock-in: The use of MCP allows the RAG server to be swapped or upgraded without rewriting the client applications (Hosts).

Security: The implementation of Milvus Partition Keys provides a blueprint for secure, multi-tenant SaaS deployments of RAG, where data isolation is a non-negotiable requirement.

Cost Control: The Semantic Cache and Context Compression features directly address the operational costs of running LLMs (token usage) and the infrastructure costs of vector search.

\subsection{Scalability and Limitations}
While the system shows promise, the reliance on LLMs for routing, rewriting, and evaluation (CRAG) introduces a dependency on the speed and cost of the LLM provider (Qwen3-8B in this case). As query volume scales, the "Complex Route" may become a bottleneck. Future work could investigate the distillation of these routing and evaluation tasks into smaller, specialized models (SLMs) to further reduce latency without sacrificing decision quality.

Additionally, the External Route relies on Tavily. In highly regulated enterprise environments where external internet access is restricted, this fallback mechanism would need to be adapted to search internal federated knowledge bases instead of the open web.

\section{Conclusion}
This report has provided an in-depth analysis of the Higress RAG MCP Server, a system designed to overcome the limitations of traditional Retrieval-Augmented Generation. By implementing a "Holistic Optimization" strategy, Higress successfully addresses the critical issues of inaccurate retrieval, hallucination, and latency in enterprise environments.

The architecture's strength lies in its integrated approach:

\begin{itemize}
    \item \textbf{Pre-Retrieval:} Adaptive Routing and HyDE ensure the system understands the user's true intent, effectively handling both simple factual queries and complex reasoning tasks.
    \item \textbf{Retrieval:} The combination of Hybrid Search (Dense + Sparse) and the \textbf{Metadata-Aware Boost Ranker} provides a robust safety net. This ensures that high-authority, up-to-date documentation is prioritized over outdated or community-generated content.
    \item \textbf{Post-Retrieval:} The CRAG Evaluator acts as a quality gatekeeper, filtering out irrelevant contexts to prevent the generation of misleading information.
    \item \textbf{Infrastructure:} The adoption of the Model Context Protocol (MCP) and Milvus Partition Keys ensures the system is standard-compliant, scalable, and secure for multi-tenant deployment.
\end{itemize}

Experimental results on proprietary Higress datasets demonstrate that these optimized strategies significantly enhance retrieval recall (exceeding 90\%) and factuality compared to naive baselines, while Semantic Caching maintains millisecond-level latency for high-frequency queries. As RAG systems continue to evolve, the modular, protocol-driven design of Higress offers a compelling blueprint for the next generation of intelligent knowledge management systems.




\end{document}